\documentclass[12pt]{article}
\usepackage{epsfig}
\newcommand{\bq}{\begin{quote}}
\newcommand{\eq}{\end{quote}}
\newcommand{\ben}{\begin{enumerate}}
\newcommand{\een}{\end{enumerate}}
\newcommand{\bi}{\begin{itemize}}
\newcommand{\ei}{\end{itemize}}
\newcommand{\ket}[1]{\vert#1\rangle}
\newcommand{\cS}{{\cal S}}
\begin{document}
\setlength{\baselineskip}{16.5pt}
\title{Defending the Pondicherry interpretation: A response to Shafiee, Jafar-Aghdami, and Golshani}
\author{U. Mohrhoff\\
Sri Aurobindo International Centre of Education\\
Pondicherry 605002 India\\
\normalsize\tt ujm@auromail.net}
\date{}
\maketitle
\begin{abstract}\noindent 
Recently Shafiee, Jafar-Aghdami, and Golshani (Studies in History and Philosophy of Modern Physics, 37, 316--329) took issue with certain aspects of the Pondicherry interpretation of quantum mechanics, especially its definitions and uses of ``objective probability,'' its conception of space, the role it assigns to the macroworld in a universe governed by quantum laws, and its claim for the completeness of quantum mechanics. These issues are addressed.

\vspace{6pt}\noindent {\it Keywords\/}: quantum mechanics; Pondicherry interpretation; objective probability; macroscopic; supervenience; space

\setlength{\baselineskip}{14pt}
\end{abstract}
\pagebreak
\section{Introduction}
A couple of years ago a critique of my interpretation of quantum mechanics---in several previous publications referred to as ``the Pondicherry interpretation of quantum mechanics'' or PIQM---was published in this journal (Shafiee, \textit{et al.}, 2006). On the positive side, Shafiee, Jafar-Aghdami, \&\ Golshani (SJG) maintain that I introduce ``a new understanding of spatiotemporal events, the character of physical reality and the meaning of objective probability,'' and that
\bq
this interpretation has some attractions. At least, some parts of the ontological attitude of Mohrhoff about the quantum world [are] fascinating. For example, his view about the supervenience of the reality of phenomena in micro-world on the events of macro-world or his metaphysical view about the `space as the totality of existing spatial relations' deserves attention. Mohrhoff's interpretation decisively affects the way one observes the quantum world.
\eq
However, believing---and rightly so---``that our understanding of the quantum phenomena should be compatible with all levels of physical description,''  SJG doubt that the PIQM ``can satisfactorily supply such compatibility.'' They perceive ``imperfections and incoherencies involved in Mohrhoff's conceptualization of space and time.'' The purpose of this article is to dispel these doubts and to explain why the perceived imperfections and incoherencies are misperceptions.

The four main concerns of SJG are my views on quantum-mechanical probabilities, space, the macroworld, and the completeness of quantum mechanics. These concerns are addressed in Secs. 2, 3, 4, and 5, respectively. The final section aims to take the mystery out of what is probably the most challenging feature of the PIQM---the supervenience of the microscopic on the macroscopic.

\section{Probability}
The PIQM is an interpretation of unadulterated, standard quantum mechanics. It is offered as an attempt to make sense of a hypothetical world in which the quantum-mechanical probability assignments are always exactly right, without nonlinear or stochastic modifications to Schr\"odinger's equation. It interprets a hypothetical world in which the reason why hidden variables are hidden is that hidden variables---be they local or nonlocal, contextual or noncontextual---do not exist. By the same token, it does not countenance absolute probabilities.%
\footnote{\label{primas}The conditionality of quantum-mechanical probabilities was also stressed by Primas (2003), who draws attention to an axiomatic alternative to Kolmogorov's (1950) formulation of probability theory, due to R\'enyi (1955, 1970), and points out that every result of Kolmogorov's theory, in which absolute probabilities have primacy over conditional ones, has a translation into R\'enyi's theory, which is based entirely on conditional probabilities.}
In addition it assumes that an arbitrarily small quantitative difference---for instance, the difference between probability~1 and a probability arbitrarily close to but less than~1, or the difference between off-diagonal matrix elements equal to~0 and off-diagonal matrix elements arbitrarily close but unequal to~0, or the difference between exact bi-orthogonality and however-near bi-orthogonality---does not account for the significant difference between the possession of a property (by a physical system) or a value (by an observable) and the lack thereof (Mohrhoff, 2004a). In other words, it interprets a hypothetical world without ``elements of reality,'' in which neither probability~1, nor the diagonality of a reduced density-matrix, nor exact bi-orthogonality is sufficient for \textit{is} or \textit{has}.%
\footnote{Another reason why probability~1 is not sufficient for \textit{is} or \textit{has} is that implicit in every quantum-mechanical probability assignment is the assumption that a measurement is successfully made: there is an outcome. (After all, this is the reason why the probabilities of the possible outcomes of a measurement add up to~1.) This also holds in the special case in which the quantum-mechanical probability of a particular outcome equals~1. Quantum mechanics therefore yields probabilities with which this or that outcome is obtained in a successful measurement, not probabilities with which this or that property or value is possessed, regardless of measurements.}

Then what is? According to the PIQM, to \textit{be} is to be \textit{measured}, and \textit{any} event or state of affairs from which the truth or falsity of a proposition of the form ``system~$S$ has the property~$P$'' (or ``observable~$O$ has the value~$V$'') can be inferred, qualifies as a measurement. No property or value is possessed unless its possession is indicated by, or inferable from, an actual event or state of affairs.%
\footnote{There is no need for anyone to actually make the inference. Nor do we need to define ``actual events'' or ``states of affairs'': we know them if we see them. Nor can we account for their occurrence or existence, inasmuch as the quantum-mechanical probability calculus presupposes it.}
The properties of quantum systems are \textit{extrinsic} in this particular sense. They \textit{supervene} on property-indicating events (in this particular sense).

The quantum laws are correlation laws: they quantify correlations between (primarily) property-indicating events and (secondarily) properties indicated by events. Since the PIQM does not countenance absolute probabilities (cf. Note \ref{primas}), the existence of correlations between measurement outcomes is tantamount to the dependence of the probabilities of possible outcomes on actual (or assumed) outcomes. If we use the quantum-mechanical correlation laws to assign probabilities,  we are free to choose (i)~the actual or assumed outcome (or outcomes) on the basis of which we assign probabilities, and (ii)~the measurement (or measurements) to the possible outcomes of which we assign probabilities. SJG are thus right in saying that I differentiate between the laws of physics (qua correlation laws) and the way we make use of them, but they are wrong in claiming that, according to me, ``probability assignments have to be considered as fundamental as the quantum laws themselves.'' I consider the correlation laws ``more fundamental'' than the probabilities we assign with their help (for once allowing ``fundamental'' to have a comparative), for whereas we cannot change the correlation laws, we can choose how we use them. The probabilities of possible measurement outcomes depend on actual (or assumed) measurement outcomes of \textit{our choice}.

I nevertheless believe that there are good reasons to consider quantum-mechanical probabilities \textit{objective}:%
\footnote{These are spelled out in greater detail in Mohrhoff (2009b), which of course was not available to SJG at the time their writing.}
\ben
\item They are assigned on the basis of (i) objective, value-indicating events or states of affairs and (ii) objective physical laws.
\item They play an essential role in the description of objective reality.
\item They are needed to define and quantify an objective indeterminacy or ``fuzziness.'' (According to the PIQM, the proper way to define and quantify a fuzzy state of affairs is to assign probabilities counterfactually, to the possible outcomes of \textit{unperformed} measurements.)
\item They do not reduce to Bayesian degrees of belief. (The stability of atomic hydrogen rests on the objective fuzziness of its internal relative position and momentum, not on anyone's belief about the values of these observables.)
\een
How could one describe a quantum system \textit{between} measurements---\textit{without} transmogrifying a probability algorithm like the wave function or the state vector into an evolving state of affairs? The PIQM stipulates that even the description of a quantum system \textit{between} property-indicating events be based on property-indicating events. Suppose, for instance, that measurements have been made at $t_1$ and~$t_2$, respectively, and that no measurements have been made in the meantime. The facts relevant to the description of the system between $t_1$ and~$t_2$ are then all of the property-indicating events---in particular, the measurements made at $t_1$ and~$t_2$---that bear upon the probabilities of the possible outcomes of all measurements that could have been made between $t_1$ and~$t_2$, but were not (Mohrhoff, 2000, 2001, 2004a, 2008). We obtain a description of the fuzzy state of affairs between $t_1$ and~$t_2$ by assigning probabilities to the possible outcomes of these measurements on the basis of those facts. Since this state of affairs, like every possessed system property, supervenes on property-indicating events, it cannot be thought of as a repository of \textit{propensities} or \textit{potentialities} existing over and above the actual property-indicating events.

Thus there are also senses in which quantum probabilities do \textit{not} qualify as objective:
\ben
\item As single-case probabilities, they are not objective in the sense of being relative frequencies.%
\footnote{When it comes to \textit{measuring} single-case probabilities, we must of course resort to relative frequencies. Nevertheless, as Popper (1982) remarked, it isn't the case that something is more likely \textit{because} it happens more often; rather, something happens more often \textit{because} it is more likely.}
\item In spite of their being single-case probabilities, they are not objective in the sense of being propensities or potentialities.
\een
In addition to giving reasons why (and why not) quantum probabilities are objective, I have argued that they have objective as well as subjective uses (Mohrhoff, 2001). One arrives at an objective description of the state of affairs that obtains between measurements if and only if, in assigning probabilities to the possible outcomes of unperformed measurements, all relevant facts are taken into account. In general this includes past as well as future events, for quantum mechanics allows us to assign probabilities on the basis not only of earlier \textit{or} later measurement outcomes using Born's rule but also of earlier \textit{and} later measurement outcomes using the ABL rule%
\footnote{The statement of the ABL rule in (Shafiee, \textit{et al.}, 2006) is incorrect. In place of the absolute \textit{values} in equation (1) there should be absolute \textit{squares}.}
(Aharonov, \textit{et al.}, 1964). If one assigns probabilities to the possible outcomes of unperformed measurements, with a view to describing a fuzzy state of affairs obtaining between measurements, without taking all relevant facts into account, one arrives at a description that is marred by a subjective element of ignorance. This subjective contamination is what makes some probability assignments subjective (in this particular sense).

The following statements by SJG are thus either liable to be misunderstood or incorrect, and if incorrect then due to either a misunderstanding on the part of SJG or an incorrect statement in one of my earlier papers. (My purpose here is to clarify rather than to sort this out.)
\bq
[According to Mohrhoff] objective probability should not merely be attributed to actually observed measurement results.
\eq
As a matter of fact, the assignment of a probability to the outcome of an \textit{actually performed} measurement is always \textit{subjective}, inasmuch as it fails to take the outcome of this measurement into account. Only probabilities assigned to the possible outcomes of \textit{unperformed} measurements can be objective.
\bq
According to Mohrhoff, what the ABL rule basically shows is that for calculating quantum probabilities both the initial and the final states of the system are to be known.
\eq
This is an odd way of saying that if one wants to assign probabilities on the basis of earlier \textit{and} later outcomes, then one has to use the ABL rule. But certainly there is no obligation to do so. One may choose to assign probabilities to possible outcomes on the basis of \textit{any} set of relevant actual outcomes. One only should distinguish between such probability assignments as take \textit{all} relevant facts into account (and therefore contribute to describe an objective fuzziness) and such as fail to do so.
\bq
One of the consequences of introducing objective probability in quantum theory is objective indefiniteness, in Mohrhoff's view.
\eq
This rather puts the cart before the horse. According to the PIQM, objective indeterminacy is a salient feature of the physical world. Assigning probabilities to the possible outcomes of unperformed measurements is merely a way---albeit the only one known to me---of describing a fuzzy state of affairs.
\bq
He considers probabilistic propositions in quantum mechanics to be counterfactual\dots\ in his view\dots\ the attribution of probability to the results of actually performed experiments is not appropriate\dots
\eq
This can hardly be my position, for if probabilities could not be attributed to the outcomes of performed measurements, quantum mechanics would not be a testable theory. Once again, while probabilities can be assigned to the possible outcomes of both performed and unperformed measurements, those assigned to performed measurements are subjective in the sense just spelled out.
\bq
[N]either correlations nor correlata have physical reality in Mohrhoff's view. 
\eq
Mermin (1998) has denied physical reality to the correlata. Nobody, to my knowledge, has denied physical reality to \textit{both} correlations and correlata. The PIQM denies physical reality only to \textit{unmeasured} properties or values. 
\bq
Marchildon (2004) challenges Mohrhoff's speculation, since for him the non-valuedness of [an unmeasured observable] does not logically follow from a counterfactual construction of the ABL rule.
\eq
Marchildon's concerns are addressed in (Mohrhoff, 2004a). Even though the valuedness of unmeasured observables is inconsistent with a variety of `no-hidden-variables' theorems (given the assumptions under which they are proved), I never claimed that the non-valuedness of unmeasured observables is a \textit{logical} consequence of anything. 
\bq
Kastner (2001) has\dots\ argued that Mohrhoff's application of the ABL rule\dots\ fail[s] to escape the conclusion of the proofs which state that the counterfactual usage of the ABL rule yields consequences that are inconsistent with quantum theory.
\eq
For refutations of these alleged ``proofs'' (Sharp \& Shanks, 1993; Cohen, 1995; Miller, 1996; Kastner, 1999ab) see (Mohrhoff, 2001) as well as (Vaidman, 1999). For a detailed response to Kastner (2003), which SJG do not cite, see (Mohrhoff, 2008).
\bq
[I]f we accept [Mohrhoff's] attitude, we must necessarily be committed to the irreducibility of the concept of probability\dots\ If the concept of probability is taken to be a non-reducible one, the realization of some counterfactual statements in the measurement process has no justification, save mere chance. But, to avoid a notion of objective chance, Mohrhoff introduces the concept of objective indefiniteness\dots
\eq
I indeed hold that possibility and its quantification, probability, are irreducible concepts. Thus I have neither reason nor the desire to avoid objective chance. Nor, therefore, do I introduce the concept of objective indeterminacy for this purpose. On the contrary, objective fuzziness is the primary concept. This gives rise to the question of how best to describe it in mathematical terms. And my answer to this question is, by assigning irreducible probabilities to the possible outcomes of (unperformed) measurements.
\bq
[O]nce one considers objective probabilities to be statistical distributions over counterfactual statements, there appears a possibility of defining joint probabilities for incompatible quantities. One can take into account a complete set of eigenvalues of two incompatible observables, and if the statistical distributions do not refer to real events, one can define joint probabilities for the results of those two quantities\dots
\eq
Even \textit{unperformed} measurements are \textit{measurements}. If two measurements are incompatible in the actual world, they are also incompatible in all nomologically possible worlds. If, for instance, the $x$ and $y$~components of a spin-$1/2$ system are measured at the times $t_1$ and~$t_2$, respectively, if both outcomes are ``up'', and if the system is not subjected to any measurement in the meantime, then the following counterfactuals are true:
\bi
\item If the $x$~component had been measured in the meantime (other things being equal), then the outcome ``up'' would have been obtained with probability~1.
\item If the $y$~component had been measured in the meantime (other things being equal), then the outcome ``up'' would have been obtained with probability~1.
\item If first the $x$~component and then the $y$~component had been measured in the meantime (other things being equal), then the outcome ``up'' would have been obtained with probability~1 in both measurements.
\item If first the $y$~component and then the $x$~component had been measured in the meantime (other things being equal), then the outcome ``up'' would have been obtained with probability~$1/2$ in both measurements.
\ei
All of these counterfactuals (and infinitely many others) contribute to describe the fuzzy state of affairs that obtains between $t_1$ and~$t_2$. What does \textit{not} contribute to its description is counterfactuals with nomologically impossible  antecedents, such as ``if both the $x$ and $y$~components had been measured \textit{simultaneously} in the meantime.''

\section{Space}
The PIQM arrives at its ontological affirmations by analyzing quantum-mechanical probability assignments in a variety of measurement contexts, rather than via an ontologization of mathematical algorithms or symbols. This is most straightforwardly done by adopting the approach popularized by Feynman (Feynman, \textit{et al.}, 1965), according to which probabilities are added if alternatives are experimentally distinguishable, whereas amplitudes are added if alternatives are experimentally indistinguishable. In other words, if one wants to calculate the probability of a particular outcome of a measurement~$M_2$, given the actual outcome of an earlier measurement~$M_1$, one must choose a sequence of measurements that may be made in the meantime, and apply the appropriate rule:
\bi
\item If the intermediate measurements are made (or if it is possible to infer from other measurements what their outcomes would have been if they had been made), one first squares the absolute values of the amplitudes associated with the alternatives and then adds the results.
\item If the intermediate measurements are not made (and if it is not possible to infer from other measurements what their outcomes would have been), one first adds the amplitudes associated with the alternatives and then squares the absolute value of the result.
\ei
The principal interpretational strategy of the PIQM is now readily stated:
\bi
\item Whenever quantum mechanics requires that amplitudes be added, the distinctions we make between the possible sequences of intermediate outcomes (``alternatives'') are distinctions that ``Nature does not make'': they correspond to nothing in the actual world; they exist solely in our minds.
\ei
The paradigm example is the two-slit experiment with electrons (Feynman, \textit{et al.}, 1965, Secs. 1.1--6). It has been said that an electron---or a fullerene, for that matter (Arndt, \textit{et al.}, 1999)---can go simultaneously through more than one slit.%
\footnote{``Since both slits are needed for the interference pattern to appear and since it is impossible to know which slit the electron passed through without destroying that pattern, one is forced to the conclusion that the electron goes through both slits at the same time'' (Encyclop\ae dia Britannica, 2006).}
The reason why, according to the PIQM, the electron can do this, is that space isn't an intrinsically differentiated expanse.%
\footnote{If space were such an expanse, an extended object would have as many parts as the space it "occupies." Conversely, the unity of an object that can simultaneously be in more than one place, entails the unity of space.}
 If the setup demands that amplitudes be added, the distinction we make between the alternatives ``the electron went through the left slit'' and ``the electron went through the right slit'' has no counterpart in the actual world. And the reason why, according to the PIQM, this distinction can be without an objective counterpart, is that the difference between the two alternatives rests on spatial distinctions that are not real \textit{per se}. There is no intrinsic partition of space that forces the electron to choose which ``region of space'' to be in. On the contrary, combined with (the interference phenomena predicted by) the second rule, the unity of the electron---or the \textit{logical} unity of the fullerene's center-of-mass---militates against the conception of space as an intrinsically partitioned expanse.

According to the PIQM, physical space is a set of relations; it contains---in the proper, set-theoretic sense of ``containment''---the spatial relations that hold among material objects. The form of a composite object is the totality of its internal spatial relations. A particle lacking internal relations is therefore a formless rather than a literally pointlike object.%
\footnote{A pointlike form would be another hidden ``variable.''}
Space thus contains the forms of all things that have forms, but it does not contain the formless so-called ``ultimate constituents of matter'' (quarks and leptons, according to the Standard Model of particle physics). It is the web spun by their (more or less fuzzy) relations. Nor is there such a thing as \textit{empty} space, not because space is teeming with virtual particles or vacuum fluctuations, but because unpossessed positions do not exist; where ``there'' is nothing, there is no \textit{there}.

Just as classical mechanics idealizes by assuming exact positions, so standard quantum mechanics idealizes by assuming that the possible outcomes of a position measurement correspond to a partition ``of space.'' Because space is not intrinsically partitioned, spatial distinctions are relative and contingent: \textit{relative} because the distinction between (what we conceive of as) two disjoint regions (e.g., inside~$R$ and outside~$R$) may be real for one object and nonexistent for another (or for the same object at a different time); and \textit{contingent} because the reality of that distinction for a given object~$O$ (at a given time~$t$) depends on whether the corresponding proposition (``$O$~is in~$R$ at~$t$'') has a definite truth value (either ``true'' or ``false''), and this in turn depends on whether a definite truth value is indicated.

A particle detector is therefore needed not only to indicate the presence of a particle in the detector's sensitive region but also to \textit{realize} (make real) the \textit{distinction} between being inside and being outside that region, thereby making it possible to attribute to a particle either property. Generally speaking, the measurement apparatus, presupposed by every quantum-mechanical probability assignment, is needed not only for the purpose of indicating the possession a particular property or value but also for the purpose of realizing a set of properties or values, which thereby become available for attribution.

SJG's presentation of these ideas is, to say the least, awkward. Particularly obfuscating is their frequent use of the word ``space'' in place of ``position,'' for example when they write that 
\bq
space could exist for an object in a definite instant, but not [exist] for another object at the same instant and not [exist] for the same object at another instant,
\eq
or that, in the context of a two-slit experiment performed under the conditions stipulated by the ``add amplitudes'' rule, one
\bq
can only talk about both the $L$ and~$R$ slits as the space possessed by the object.
\eq
SJG argue that the detection of a single electron at the backdrop 
\bq
does not say anything about the position of the electron in the whole $L$ and $R$ region\dots.
[T]he statement `The electron's position is in the entire region $L\&R$' is not true, because this statement cannot be justified by the indication of an \textit{individual} particle.
\eq
As a matter of fact, it can. In the experiment under consideration, there is a single source of free electrons in front of the slit plate. The detection of a single electron at the backdrop therefore not only indicates a position at the backdrop but also that this electron went through the slits (inasmuch as with both slits shut, no electron reaches the backdrop). According to the interpretational strategy adopted by the PIQM, it is therefore true that the electron went through $L\&R$, whereas under the conditions stipulated by the ``add amplitudes rule'' neither ``the electron went through~$L$'' nor ``the electron went through~$R$'' is true. As far as the electron is concerned, $L\&R$ is an undifferentiated whole.

Nor can SJG justify their claim that the statement they quote is ``not consistent with the description given by quantum mechanics.'' What description is given by quantum mechanics is, to say the least, moot. The PIQM may of course be inconsistent with whatever interpretational scheme SJG  have at the back of their minds. Conceding the conceivability of that statement \textit{provided} that $L\&R$ refers to the electron's ``probability space'' rather than to ``a real space,'' SJG write  
\bq
Mohrhoff tries to identify the probability space with the real space, but this interpretation is not compatible with what quantum mechanics describes.
\eq
There is no question of my identifying ``probability space'' (presumably, a set of potentially attributable regions) with ``the real space.'' As said, the PIQM views physical space as made up, not of regions, but of the spatial relations (relative positions and orientations) that hold among material objects. At the same time it accepts the standard quantum-mechanical representation of the possible outcomes of a position measurement---which always measures a \textit{relative} position---as a partition ``of space'' (i.e., in SJG's terms, a subset of probability space). In fact I insist that these spaces be carefully distinguished.

SJG proceed to consider two EPR-entangled particles that are separated albeit not in a spacelike manner: ``the separation of [the] two particles could be taken to be such that they could have local interaction.'' The mistake here is that the property of being (or not being) separated in a spacelike manner can only be attributed to events; it cannot be attributed to particles. Now suppose that particle~1 is subjected to a measurement, while in the vicinity of particle~2 there is no apparatus capable of performing a measurement.
\bq
Nevertheless, there might be a local causal link between the two particles\dots. If there could be a local causal link between the two correlated particles, while the physical properties of only one of them is indicated, what can one say about the physical properties of the other one? According to quantum mechanics, in such circumstances, the whole system is described by an entangled state.
\eq
I suppose that by a ``local causal link'' the authors mean a locally mediated causal link. But what has entanglement to do with the existence or nonexistence of a spacelike relation? Nothing at all. 
\bq
Once one measures a correlated property for one of the particles, the same property is also measured indirectly for the other particle. Therefore in our proposed experiment, a measurement of, e.g., the first particle's position implies a measurement for the position of the second particle and vice versa.
\eq
This may well be the authors' opinion, as it seems to have been the opinion of Einstein, \textit{et al.} (1935), but it certainly isn't part of unadulterated quantum mechanics, which consists of (i)~a calculus that yields conditional probabilities or correlation laws and (ii)~assumptions about the nature of the correlata, such as those dubbed by Redhead (1987) the ``minimal instrumentalist interpretation of quantum mechanics.'' SJG then notice 
\bq
a gap between Mohrhoff's interpretation and a minimal causal description of quantum mechanics [which] discriminates between the two notions of indication and measurement (at least, in von Neumann's approach who describes the measurement as an interaction followed by a collapse)\dots
\eq
Anyone modestly familiar with the PIQM will be aware that it repudiates von Neumann's approach, which therefore cannot be invoked for the purpose of ``demonstrating the deficiency'' of my reasoning. The PIQM rejects not merely the notion of collapse but the very notion that is responsible for the mother of all quantum-mechanical pseudo-problems---the notion of quantum state \textit{evolution}. A quantum state $\ket{\psi(t)}$ is not an evolving, instantaneous state of affairs that obtains at the time~$t$, and that collapses (or appears to collapse) at the time of a measurement. It is an algorithm that serves to assign probabilities to the possible outcomes of any measurement that may be performed---either actually or counterfactually---at the time~$t$. The parameter~$t$ refers to the time of this measurement. Without reference to a measurement, it is ill-defined.

As for the gap between the PIQM and a minimal causal interpretation of quantum mechanics, that certainly exists. The fact that value-indicating events lack causally sufficient conditions (Ulfbeck \&\ Bohr, 2001; Mohrhoff, 2002b, 2004a) alone is sufficient to argue that the quantum-mechanical correlations between such events do not admit of a causal interpretation. In my opinion, the search for a causal explanation of these correlations puts the cart before the horse. It is the laws governing these correlations that determine the extent to which causal concepts can be used. Such concepts are applicable only to the macroworld (see below), where the statistical correlation laws of quantum physics degenerate into the deterministic laws of classical physics, for this alone makes it possible to think of the correlata in terms of causes and effects.

\section{The macroworld}
SJG maintain that both the Copenhagen interpretation and the PIQM ``require something beyond a quantum process for attributing physical reality to the realm of micro-physics.'' While I cannot vouch for the Copenhagen interpretation, which comes in a variety of flavors, there is no such thing as a quantum process where the PIQM is concerned. It rejects the notion of a quantum process for the same reason that it rejects the notion of quantum state evolution. What there is, according to it, is property-indicating events, properties indicated by events, and the \textit{macroworld}, which encompasses the property-indicating events as unpredictable changes in the values of macroscopic positions (which is short for ``relative positions between macroscopic objects''). Whereas the indicated properties owe their existence to the indicating events, the macroworld is a completely self-contained system of relations; it depends on nothing external to itself.%
\footnote{The arguments leading to this conclusion can be found several of my papers (Mohrhoff, 2002b, 2004b, 2005, 2006, 2009bc) and need not be recapitulated here.}
SJG nevertheless detect a ``hidden role of observer in Mohrhoff's interpretation'':
\bq
one could declare that an indication without an observer has no meaning in the quantum domain, even if the classical objects are considered to be self-indicating\dots\ For without an observer, it cannot be verified if any property is really indicated for quantum systems.
\eq
As said, the PIQM views property-indicating events as unpredictable changes in the values of macroscopic positions. Such changes take place irrespective of whether anyone is around to observe them.

Whereas every sufficiently comprehensive physical theory defines a set of nomologically possible worlds, no physical theory can differentiate between the actual world and another nomologically possible world (let alone account for the existence of the actual world). In classical physics we identify the actual world as the possible world whose initial (or initial and final) conditions match the \textit{observed} initial (or initial and final) conditions. In quantum physics we identify the actual world as the possible world whose property-indicating events match the \textit{observed} property-indicating events. If the conclusion that classical physics presupposes conscious observers is unwarranted, as most everyone agrees, then the conclusion that quantum physics presupposes conscious observers is equally unwarranted.

In reality, we are not given a set of nomologically possible worlds and required to identify the actual world. We are given the actual world, whose laws define a set of possible worlds. The question is, do these laws admit the construction of a theoretical model that can be thought of as the model of a free-standing (Fuchs \&\ Peres, 2000ab) or strongly objective (d'Espagnat, 1989, 1995) reality capable of existing \textit{without} being given? In other words, can the ``full elision of the subject'' (Bitbol, 1990) be achieved? As far as classical physics is concerned, the widely accepted answer is Yes. According to the PIQM, the answer is equally affirmative where quantum physics is concerned. It is, however, less straightforward to arrive at. One has to decide which of the structures that can be constructed on the foundation of the mathematical formalism and its minimal instrumentalist interpretation (Redhead, 1987), describes what is independently real. There is no doubt in my mind that this can only be the macroworld.

According to SJG, 
\bq
we are always confronted with the question of how can one be assured of whether a measuring apparatus performs a measurement or not? For example, in Schr\"odinger's cat experiment, one cannot distinguish that the cat is dead or alive\dots\ when the cat is not observed and there is no perception of what was being measured\dots
\eq
This is the kind pseudo-question that arises if one follows von Neumann in transmogrifying a probability algorithm that depends on the time of a measurement into an evolving state of affair. We are not given a set of possible measurement outcomes and required to identify the actual outcome. We are given actual outcomes and laws that govern their correlations. We are given cats that are either dead or alive. The difference between dead cats and living cats involves macroscopic differences that do not require confirmation by observers.

SJG also miss ``a quantitative criterion for defining the sharpness of a spatial distribution.'' I wonder why, since quantitative measures for the spread of probability distributions are common in standard probability theory. SJG further point out that we may not be able to distinguish between a sharp distribution and a non-sharp distribution. I agree, but the PIQM crucially defines ``macroscopic'' in a way that does not depend on this distinction. 

Again, ``could one say,'' SJG ask,
\bq
that in the absence of the classical world there was no room for anything else to be in existence?
We should not forget that we did not have a world obeying the rules of classical mechanics from the beginning. The existence of [a] classical world, with objects having sharp spatial distribution, requires certain conditions which have not been there all the time. For example, what kind of reality could one ascribe to fundamental particles in the early universe, when there was no indication? Apparently, Mohrhoff takes it for granted that the classical world has always been there and that it will exist in the future.
\eq
These are intriguing questions, to which I hazarded answers in a paper that SJG seem to have missed (Mohrhoff, 2002d). Once again, according to the PIQM, to \textit{be} is to be \textit{measured}. The microscopic \textit{supervenes} on the macroscopic. What happens or is the case in the microworld depends on the goings-on in the macroworld, rather than the other way around, as we are wont to think. Much the same applies to the pre-macroscopic era. Like the microworld, this supervenes on the macroworld. The macroworld does \textit{not} extend infinitely into the past (nor, presumably, into the future), but just as microscopic events are indicated by macroscopic events, so are pre-macroscopic events. The reality that one can ascribe to the properties of the universe at pre-macroscopic times is of the same kind as the reality that one can ascribe to the properties of a quantum system~$\cS$ between measurements, except that what obtains \textit{between} measurements depends on probability assignments based on earlier \textit{and} later events via the ABL rule, whereas what obtains \textit{before} the macroscopic era depends on probability assignments based on later events via the Born rule.%
\footnote{Thus the only density operator that one can meaningfully associate with the pre-macroscopic universe is an advanced or ``retropared'' one---a density operator that ``evolves'' toward the past in the same (spurious) sense in which a retarded or ``prepared'' density operator ``evolves'' toward the future.}

\section{Completeness}
It is generally taken for granted that if quantum states are probability algorithms, then quantum mechanics is an incomplete theory.%
\footnote{For instance: ``The quantum wave function $\psi$ might be merely a mathematical tool for calculating and predicting the measured frequencies of outcomes over an ensemble of similar experiments\dots\ However, even in the statistical interpretation, the `measurement problem' in the more general sense remains. For quantum theory is then an incomplete theory that refers only to ensembles'' (Bacciagaluppi and Valentini, 2006).}
This conclusion would be warranted if quantum-mechanical probabilities were defined as relative frequencies (which according to the PIQM is not the case) or if quantum mechanics were incapable of encompassing the value-indicating events to which it serves to assign probabilities (which, as we have seen, is also not the case). What about SJG's objection that in ``all interpretations that consider quantum mechanics to be complete and final\dots\ there are always certain questions that physics should not attempt to answer, as they go beyond the quantum description''?

No matter what the fundamental theory eventually turns out to be, as long as there is a fundamental theory, there will always be the mystery of its origin. A fundamental theory will always be incomplete in this minimal sense.%
\footnote{ If at all a fundamental theory can be explained, it is in weakly teleological terms. For quantum theory, such an explanation has been offered in (Mohrhoff, 2009a). For the Standard Model, it has been outlined in (Mohrhoff, 2002c).}
It will also be incomplete in the sense that it cannot explain why there is anything, rather than nothing at all.

As it is the objective of the PIQM to make sense of a hypothetical world in which the quantum-mechanical probability assignments are always exactly right, I take it for granted that the fundamental theory is quantum-mechanical in nature. If there are additional questions that quantum mechanics fails to answer, it is (I believe) not because the theory is incomplete but because \textit{the physical world} is incomplete---as compared to certain theoretical expectations that have psychological underpinnings (Mohrhoff, 2006, 2007) but are physically unwarranted.
\bi
\item Whereas the quantum-mechanical probability algorithm cannot provide sufficient conditions for the occurrence of value-indicating events, this signals the incompleteness of the theory only if such conditions nevertheless exist; otherwise it signals an ``incompleteness'' of the physical world.
\item Whereas the use of causal concepts is confined to the macroworld, this signals the incompleteness of the theory only if a micro causal nexus nevertheless exists; otherwise it signals another ``incompleteness'' of the physical world.
\item Whereas no values can be assigned to unmeasured observables, this signals the incompleteness of the theory only if unmeasured observables nevertheless have values; otherwise it signals yet another ``incompleteness'' of the physical world.
\ei
In fact, the incomplete differentiation of the physical world (spatiotemporal as well as substantial) is (i)~a straightforward consequence of the interpretational strategy of the PIQM and (ii)~one of its most significant implications. If in our minds we partition the world into smaller and smaller spatial regions, there comes a point beyond which there is no material object for which these regions, or the corresponding distinctions, exist. \textit{Mutatis mutandis}, the same holds for the temporal and substantial distinctions we make (Mohrhoff, 2005, 2006, 2009bc). They are warranted by property-indicating events, and the latter do not license an absolute and unlimited objectification of the former.

\section{Outlook}
According to Schr\"odinger (1935), entanglement is ``not\dots\ one but rather \textit{the} characteristic trait of quantum mechanics.'' According to Misner \textit{et al.} (1973), the central mystery of physics is the ``miraculous identity'' (p.~1215) of particles of the same type. According to Feynman \textit{et al.} (1965, Sec.~1--1), the double-slit experiment with electrons ``has in it the heart of quantum mechanics.'' According to Stapp (1975),  Bell's theorem is ``the most profound discovery in science.'' To my mind, all of these extraordinary features of quantum mechanics are subsumed and eclipsed by the supervenience of the microscopic on the macroscopic, which flies in the face of a twenty-five centuries old paradigm. It no longer is appropriate to ask: what are the ultimate building blocks, and how do they interact and combine?%
\footnote{This follows not only from the extrinsic nature of the values of observables but also from the incompleteness of the world's substantial differentiation (Mohrhoff, 2005, 2006, 2009bc).}
Nor is the incomplete spatiotemporal differentiation of the physical world consistent with theoretical models that construct physical reality on the foundation of an intrinsically and completely differentiated spatiotemporal expanse, by associating physical properties with spacetime points. 

According to the identity of indiscernibles, $A$ and~$B$ are one and the same thing just in case there is no difference between $A$ and~$B$. There is no difference between two fundamental particles if each is considered by itself, out of relation to any other object.%
\footnote{A fundamental particle considered by itself lacks internal structure and, hence, a form. Since motion is relative, we cannot attribute to it any of the properties that derive their meanings from the quantum-mechanical description of motion (that is, from external symmetry operations). Nor can we attribute to it any kind of charge, since charges derive their meanings from the quantum-mechanical description of interactions (that is, from internal symmetry operations). Quantum statistics, finally, rules out the association of distinct identities with objects lacking persistent and unswappable properties.}
Hence, considered out of relation to their relations, all fundamental particles are identical in the strong sense of numerical identity. In a well-defined sense, therefore, the number of ``ultimate constituents'' equals one. A similar conclusion can be reached by noting that the number of particles in a relativistic quantum system is a quantum observable. As such it is a property of the system as a whole, and it has a (definite) value only if (and only when) it is actually measured. In a well-defined sense, therefore, a relativistic quantum system is an intrinsically undivided whole, including the largest conceivable system---the physical universe.

The appropriate question to ask is: how does this Ultimate Constituent \textit{manifest} itself? How does it take on properties? How does it constitute an apparent multitude of objects, and how does it realize their forms? Quantum mechanics (as seen through the eyes of the PIQM) suggest this simple answer: by entering into spatial relations with itself, the Ultimate Constituent gives rise to both matter and space, inasmuch as space is the totality of existing spatial relations, whereas matter is the corresponding (apparent) multitude of relata---``apparent'' because the relations are \textit{self}-relations.

If we equate the manifested world with the macroworld, then quantum mechanics affords us a glimpse ``behind'' the manifested world at formless particles and non-visualizable atoms, which, instead of being the world's constituent parts or structures, are instrumental in its manifestation. However, it allows us to describe what we ``see'' only in terms of inferences from macroevents and their quantum-mechanical correlations. If we experience something the like of which we never experienced before, we are obliged to describe it in terms of familiar experiences. By the same token, what lies ``behind'' the manifested world can only be described in terms of the finished product---the macroworld. I believe that this way of thinking makes the supervenience of the microscopic on the macroscopic a tad less mysterious.

%++++++++++++++++++++++++++++++++++++++++++++++++++++++++++++++++
\vspace{20pt}\setlength{\parskip}{5pt}\noindent\textbf{References}

\vspace{5pt}\leftskip=15pt\setlength{\parindent}{-15pt}
Aharonov, Y., Bergmann, P. G., \&\ Lebowitz, J. L. (1964). Time symmetry in the quantum process of 
measurement. \textit{Physical Review}, \textit{134}, B1410--B1416. 
 
Arndt, M., Nairz, O., Vos-Andreae, J., Keller, C., van der Zouw, G., \&\ Zeilinger, A. (1999). Wave-particle duality of $C_{60}$ molecules. \textit{Nature}, \textit{401}, 680--682.

Bacciagaluppi, G., \&\ Valentini, A. (2006). Draft of \textit{Quantum Theory at the Crossroads: Reconsidering the 1927 Solvay Conference}. Cambridge: Cambridge University Press (to be published); \textit{arXiv.org}, quant-ph/0609184, p. 164--165.

Bitbol, M. (1990). l'\'Elision. Preface to E. Schr\"odinger, \textit{L'esprit et la mati\`ere}. Paris: Seuil.

Cohen, O. (1995). Pre- and postselected quantum systems, counterfactual measurements, and consistent histories, \textit{Physical Review A}, \textit{51}, 4373--4380.

d'Espagnat, B. (1989). \textit{Reality and the Physicist: Knowledge, Duration and the Quantum World}. Cambridge: Cambridge University Press.

d'Espagnat, B. (1995). \textit{Veiled Reality, an Analysis of Present-Day Quantum Mechanical Concepts}. Reading, MA: Addison-Wesley.

Encyclop\ae dia Britannica (2006). Quantum mechanics. Retrieved from \textit{Encyclop\ae dia Britannica Online}: http://www.britannica.com/eb/article-77520.

Einstein, A., Podolsky, B., \&\ Rosen, N. (1935). Can quantum-mechanical description of physical reality be considered complete? \textit{Physical Review}, \textit{47}, 777--780 

Feynman, R. P., Leighton, R. B., \&\ Sands, M. (1965). \textit{The Feynman Lectures in Physics, Vol.~3}. Reading, MA: Addison-Wesley.

Fuchs, C. A., \&\ Peres, A. (2000a). Quantum theory needs no `interpretation'. \textit{Physics Today}, \textit{53} (March), 70--71.

Fuchs, C. A., \&\ Peres, A. (2000b). Quantum theory---interpretation, formulation, inspiration: 
Fuchs and Peres reply. \textit{Physics Today}, \textit{53} (September), 14, 90.

Kastner, R. E. (1999a). Time-symmetrized quantum theory, counterfactuals, and ``advanced action''. \textit{Studies in History and Philosophy of Modern Physics}, \textit{30}, 237--259.

Kastner, R. E. (1999b). The three-box ``paradox'' and other reasons to reject the counterfactual usage of the ABL rule. \textit{Foundations of Physics}, \textit{29}, 851--863.

Kastner, R. E. (2001). Comment on ``What quantum mechanics is trying to tell us'', by Ulrich Mohrhoff. \textit{American Journal of Physics}, \textit{69}, 860--863.

Kastner, R. E. (2003). The nature of the controversy over time-symmetric quantum counterfactuals. \textit{Philosophy of Science}, \textit{70}, 145--163.

Kolmogorov, A. N. (1950). \textit{Foundations of the Theory of Probability}. New York: Chelsea.

Marchildon, L. (2004). Remarks on Mohrhoff's interpretation of quantum mechanics. \textit{Foundations of Physics}, \textit{34}, 59--73. 

Mermin, N. D. (1998). What is quantum mechanics trying to tell us? \textit{American Journal of Physics}, \textit{66}, 753--767.

Miller, D. J. (1996). Realism and time symmetry in quantum mechanics, \textit{Physics Letters A}, \textit{222}, 31--36.

Misner, C. W., Thorne, K. S., \&\ Wheeler, J. A. (1973). \textit{Gravitation\/}. San Francisco: Freeman.

Mohrhoff, U. (2000). What quantum mechanics is trying to tell us. \textit{American Journal of Physics}, \textit{68}, 728--745. 

Mohrhoff, U. (2001). Objective probabilities, quantum counterfactuals, and the ABL rule: A response to R.~E. Kastner. \textit{American Journal of Physics}, \textit{69}, 864--873.
 
Mohrhoff, U. (2002a). The world according to quantum mechanics (or the 18 errors of Henry P. Stapp). \textit{Foundations of Physics}, \textit{32}, 217--254.

Mohrhoff, U. (2002b). Making sense of a world of clicks. \textit{Foundations of Physics}, \textit{32}, 1295--1311.

Mohrhoff, U. (2002c). Why the laws of physics are just so. \textit{Foundations of Physics}, \textit{32}, 1313--1324.

Mohrhoff, U. (2002d). Reflections on the spatiotemporal aspects of the quantum world. \textit{Modern Physics Letters A}, \textit{17}, 1107--1122.

Mohrhoff, U. (2004a). Do quantum states evolve? Apropos of Marchildon's remarks. \textit{Foundations of Physics,} \textit{34}, 75--97.

Mohrhoff, U. (2004b). This elusive objective existence. \textit{International Journal of Quantum Information}, \textit{2}, 201--220.

Mohrhoff, U. (2005). The Pondicherry interpretation of quantum mechanics: An over\-view. \textit{PRAMANA--Journal of Physics}, \textit{64}, 171--185.

Mohrhoff, U. (2006). Is the end in sight for theoretical pseudophysics? In V. Krasnoholovets \&\ F. Columbus (Eds.), \textit{New Topics in Quantum Physics Research}. New York: Nova Publishers.

Mohrhoff, U. (2007). The quantum world, the mind, and the cookie cutter paradigm. \textit{AntiMatters}, \textit{1}, 55--90.

Mohrhoff, U. (2008). The utility of time-symmetric quantum counterfactuals: A response to Kastner. arXiv:0801.2215v2 [quant-ph].

Mohrhoff, U. (2009a). Quantum mechanics explained. To appear in \textit{International Journal of Quantum Information}, \textit{7} (1).

Mohrhoff, U. (2009b). Objective probability and quantum fuzziness. To appear in \textit{Foundations of Physics}.

Mohrhoff, U. (2009c). A fuzzy world. In I. Licata and A. Sakaji  (Eds.), \textit{The Nature Description in Quantum Field Theory}. New York: Springer. 

Popper, K. R. (1982). \textit{The Open Universe: An Argument for Indeterminism}. London: Hutchinson.

Primas, H. (2003). Time-entanglement between mind and matter. \textit{Mind and Matter}, \textit{1}, 81--119.

Redhead, M. (1987). {\it Incompleteness, Nonlocality and Realism\/}. Oxford: Clarendon.

R\'enyi, A. (1955). A new axiomatic theory of probability.  \textit{Acta Mathematica Academia Scientiarum Hungaricae},~\textit{6}, 285--335.

R\'enyi, A. (1970). \textit{Foundations of Probability (Chap.~2)}. San Francisco: Holden-Day. 

Schr\"odinger, E. (1935). Discussion of probability relations between separated systems. \textit{Proceedings of the Cambridge Philosophical Society}, \textit{31}, 555--563.

Shafiee, A., Jafar-Aghdami, M., \&\ Golshani, M. (2006). A critique of Mohrhoff's interpretation of quantum mechanics. \textit{Studies in History and Philosophy of Modern Physics}, \textit{37}, 316--329.

Sharp, W. D., \& Shanks, N. (1993). The rise and fall of time-symmetrized quantum mechanics. \textit{Philosophy of Science}, \textit{60}, 488--499. 

Stapp, H. P. (1975). Bell's theorem and world process. \textit{Il Nuovo Cimento B}, \textit{29}, 270--276.

Ulfbeck, O., \&\ Bohr, A. (2001). Genuine fortuitousness. Where did that click come from? \textit{Foundations of Physics}, \textit{31}, 757--774.

Vaidman, L. (1999).  Defending time-symmetrized quantum counterfactuals, \textit{Studies in History and Philosophy of Modern Physics}, \textit{30}, 373--397.

\end{document}